\documentclass[aip,10pt,jcp,twocolumn,
superscriptaddress ,
showpacs,floats,floatfix,preprintnumbers,amsmath,
amssymb]{revtex4-1}
\usepackage{subfigure, graphicx}

\usepackage{array}
\usepackage{dcolumn}
\usepackage{bm}
\usepackage{booktabs}

\begin{document}

\preprint{(to be submitted to J. Chem. Phys. )}

\title{Efficient moves for global geometry optimization methods and their application to binary systems}

\author{Michael Sicher}
\affiliation{Department of Physics, Universit\"{a}t Basel, Klingelbergstr. 82, 4056 Basel, Switzerland}
\author{Stephan Mohr}
\affiliation{Department of Physics, Universit\"{a}t Basel, Klingelbergstr. 82, 4056 Basel, Switzerland}
\author{Stefan Goedecker}
\affiliation{Department of Physics, Universit\"{a}t Basel, Klingelbergstr. 82, 4056 Basel, Switzerland}

\date{\today}

\begin{abstract}
We show that molecular dynamics based moves in the Minima Hopping (MH) method are more efficient than 
saddle point crossing moves which select the lowest possible saddle point. For binary systems 
we incorporate identity exchange moves in a way that allows to avoid the generation of high energy configurations.
Using this modified Minima Hopping method we reexamine the binary Lennard Jones (BLJ) benchmark system 
with up to 100 atoms and we find a large number of new putative global minima structures. 
\end{abstract}


\maketitle

\section{Introduction}
In a global geometry optimization one has to search over many local minima until one finds the global minimum. 
Moves~\cite{ferrando_1,ferrando_2} are necessary to jump from the catchment basin of the current local minimum into another catchment basin.
The types of moves that are chosen for this hopping from one catchment basin into another has an important effect on the 
efficiency of the method. The majority of the moves used by researches in this field, fall into the following five 
categories. 
\begin{itemize}
\item Random Moves: The atoms are displaced randomly from the current positions. If the amplitude of the random displacement is large 
      enough, the system will relax into another local minimum during a standard local geometry optimization. Random moves are 
      widely used both within random search methods~\cite{needs, crystalprediction}  and within basin hopping~\cite{bashop}.
\item Molecular dynamics moves: One of the oldest global optimization methods, simulated annealing~\cite{simann} is based on a 
      modified molecular dynamics scheme where the temperature is lowered continuously. Molecular dynamics is also used as a move in the 
      Minima Hopping method~\cite{minhop} as well as in various other schemes~\cite{rogan,hennig}. 
\item Library based moves: For systems such as proteins, where one knows in advance which kind of moves are important, non-physical 
      moves, based on a library of moves can be used~\cite{librarymoves}.
\item Transition state search based moves: Starting
from a given local minimum one searches for a
neighbouring saddle point. 
      The system is then moved over the saddle point and afterwards relaxed into the local minimum across this saddle point. 
      This kind of move is used in the ART method~\cite{ART}, which is primarily a method to explore potential energy landscapes, but 
      which will also visit the global minimum if run long enough. Saddle points are also determined in the TAD method~\cite{voter} 
      which allows to determine rates of rare events. 
\item Genetic algorithms~\cite{hartke} use mutation and crossover operations which can also be considered as moves. 
      These moves are typically strongly system dependent and the moves for clusters~\cite{ho} are for instance different to the moves 
      used for crystal structure prediction~\cite{oganov}.
\end{itemize}

The density of configurations is increasing exponentially with increasing energy. It is therefore important to search 
only over an energy interval above (and including) the global minimum which is not too large. Otherwise the number of 
configurations in this interval is so large that one will miss most likely the global minimum. This means that one 
should use moves that will never or only rarely lead into high energy structures. If moves with this property are not 
used, the majority of the configurations has to be discarded in an acceptance/rejectance step which can be 
found in most global optimization algorithms. A large rejectance ratio is however also highly inefficient. 
If one considers moves that lead from 
one minimum into a neighbouring one, it has been shown
that molecular dynamics based moves are very
efficient~\cite{MDbep}. 
Since a molecular dynamics trajectory has a fixed and limited kinetic energy it follows from energy conservation that it cannot 
go over barriers that are higher than this kinetic energy. Since the Bell Evans Polanyi principle~\cite{jensen} tells us 
that minima behind low energy barriers are on average also low in energy, molecular dynamics based moves do not lead 
into high energy structures if their kinetic energy is chosen such that they can only overcome low energy barriers.

In this paper we will first compare two classes of moves that both are able to find low energy escape paths from a current 
minimum, namely molecular dynamics based moves with moves that are saddle point based. In the second part of the paper we will 
discuss moves for binary systems. We will in particular discuss under which circumstances moves that exchange the identity of two 
atoms are efficient. In the third part we will apply our resulting global optimization scheme to the benchmark set of
binary Lennard Jones (BLJ) clusters~\cite{database} and show that many global minimum structures had been overlooked in previous studies. 
The LJ potential poses the same kind of problems for global geometry optimization methods as other more realistic potentials~\cite{ferrando_3} and 
the efficiency of an algorithm for an LJ system is therefore indicative of the success for other potentials.

\section{Saddle point escape moves versus molecular dynamics escape moves}
We will now compare the efficiency of various moves within the Minima Hopping method. The original MH method consists 
of a sequence of short molecular dynamics trajectories followed by local geometry optimizations~\cite{minhop}. 
The molecular dynamics trajectory allows 
to cross barriers to hop from one catchment basin into another and the local geometry optimization will then bring the 
system to the bottom of the catchment basin.

The other kind of moves are based on saddle point searches. Starting from a local minimum, the system is propagated towards a saddle point.
After reaching such a transition state and barely crossing it, a local geometry optimization brings the system again down to a new minimum.
We use the dimer method~\cite{Henkelman} with a few modifications to search the saddle points and to identify the transition states in this modified
version of the MH method.

The dimer consists of two points $\mathbf{R}_1$ and $\mathbf{R}_2$ in the high dimensional space in which one 
wants to locate the transition state. If the dimer
midpoint is labelled as $\mathbf{R}$, then these 
two points are formed according to $\mathbf{R}_1=\mathbf{R}+\Delta R\mathbf{\hat{N}}$ and 
$\mathbf{R}_2=\mathbf{R}-\Delta R\mathbf{\hat{N}}$, where $\mathbf{\hat{N}}$ is the normalized dimer direction. 
The dimer method consists of basically two steps. In the first one, the dimer is rotated into a position 
which gives a small torsional force and which aligns it with an eigenmode of the Hessian. This torsional force is given by 
$\mathbf{F}^\perp=(\mathbf{F}_1-\bf{F}_2)-\langle\mathbf{F}_1-\mathbf{F}_2|\mathbf{\hat{N}}\rangle\mathbf{\hat{N}}$, 
with $\mathbf{F}_1$ and $\mathbf{F}_2$ being the forces at $\mathbf{R}_1$ and $\mathbf{R}_2$, respectively. 
Since the dimer midpoint remains constant during the rotation, the force acting on $\mathbf{R}_2$ can be 
approximated by $\mathbf{F}_2=2(\mathbf{F}-\mathbf{F}_1)$ in order to reduce the total number of force 
evaluations, $\mathbf{F} $ denoting the force at the dimer midpoint.
In the second step the dimer is translated along $\mathbf{F}_{eff}$, where this modified 
force is given by $\mathbf{F}_{eff}=-\langle\bf{F}|\mathbf{\hat{N}}\rangle\mathbf{\hat{N}}$ if the curvature 
along the dimer axis is negative and by $\mathbf{F}_{eff}=\mathbf{F}-10\langle\mathbf{F}|\mathbf{\hat{N}}\rangle\mathbf{\hat{N}}$ otherwise.

Whereas the translation is always done in a straightforward way, there exist several different ways how the 
rotation can be carried out. Since the rotation part requires considerably more force evaluations 
than the translation part, choosing a good rotation method is important. Most methods have the tendency 
to align the dimer along the lowest curvature mode. This is due to the fact that only the lowest 
curvature mode is a minimum with respect to the curvature; all other low curvature modes are saddle points.  
Whereas this circumstance does not cause any problem if one 
is interested in finding only one saddle point leading out of a given minimum, it is a severe 
shortcoming in our case. Due to this behaviour it is
very likely that the dimer aligns itself with 
the lowest curvature mode after a few iterations even if it was initially aligned along another 
than the lowest mode and, as a consequence, several searches will lead to the same saddle point, 
even if they were originally started in different orthogonal directions. This is a serious problem 
since we are interested in finding many different saddle points leading out of a given minimum and 
need therefore a rotation method which keeps the dimer on the initially selected mode, thus leading to distinct saddle points.

One possibility that was already mentioned in the original paper~\cite{Henkelman} is to impose the restriction 
that the dimer is orthogonal to all previous dimer directions at the minimum until the dimer is 
aligned along the lowest mode itself -- which will happen as we approach the saddle point -- since 
then there is no more tendency for the dimer to switch to another mode. This procedure requires 
however the knowledge of all lower lying modes if one is interested to follow a higher mode. 
Since in our new method we are interested in finding systematically all low lying saddle points 
around a local minimum this condition is automatically fulfilled. However, we also develop another 
method where it is necessary to directly follow a higher mode and this orthogonalization procedure 
is as a consequence not suited, and we therefore will look for another way to stay on the initially selected mode.

Tests with several rotation methods show that using Direct Inversion in Iterative Subspace (DIIS)~\cite{diis} 
is most suitable for our purpose, since DIIS has the tendency to catch the nearest lying stationary point, 
regardless of whether it is a minimum, maximum or saddle point. This allows us to stay on the initially 
selected mode with high reliability. Approximating the error vectors by $-\alpha\bf{F}^\perp$, where $\alpha$ is a constant,  
we move $\mathbf{R}_1$ according to the standard DIIS procedure with the modification that 
$\mathbf{R}_1$ has to be adjusted after each step to retain the fixed dimer separation. However, it 
turns out that it is for reasons of stability not good to stay on the initially selected mode at 
any cost, but to follow the lowest mode at some point instead. This can easily be achieved by 
abandoning the DIIS rotation and using the Lanczos method thenceforth. This switch to the Lanczos 
method was done as soon as the second derivative of the energy with respect to the number of 
iterations became negative.

In the present implementation of our saddle point searches we put the focus on reliability and not on speed, since we are only interested in understanding the principle of the various types of moves. Therefore about 1000 force evaluations are required if we want to have a success rate 
of some 99 percent. Further tuning might still bring down the number of force evaluations, but it seems unlikely that it can be reduced by one order of magnitude, which would be necessary to compete with molecular dynamics based moves. So it is clear that saddle point based moves are only of interest in practice if the global minimum can be found much faster with respect to the number of minima that have to be visited until the global one is found.

With the exception of the moves, whose details will be explained below, the standard MH algorithm 
was used, i.e. new minima are accepted if they are not higher 
than $E_{diff}$ in energy and the value of $E_{diff}$ is adjusted by a feedback mechanism such that on average half of the 
new configurations are accepted. 

We use the Lennard-Jones clusters $LJ_{55}$ and $LJ_{38}$ as test systems because they behave very differently. The  $LJ_{55}$ is a structure seeker 
for which it is very easy to find the global minimum. $LJ_{38}$ on the other hand is a two funnel 
system for which it is surprisingly difficult 
to find its global minimum in view of its small size. 100 global optimization runs are done in all cases to get well 
defined average values.

\subsection{Crossing the lowest barrier}
The first saddle point based type of move is conceptually simple. According to the Bell Evans Polanyi principle, the ideal move would be a move which escapes from the current catchment basin 
by going over the lowest barrier. There are however two problems using this type of move. In case a minimum is visited 
a second time, the sequence of minima that are visited would repeat itself ad infinitum and no new minima would be visited.
The system would not be ergodic in a certain sense. This problem can easily be overcome by a small modification of the method.
If a minimum is visited for the first time one escapes over the lowest barrier, if it is visited a second time one escapes over the 
second lowest barrier and so on. The second problem is that such a type of move would be incredibly expensive numerically. 
Doing a one sided search for a single saddle point requires typically already a few hundred force evaluations 
and exploring more or less all the saddle points around a local minimum to find out which is the lowest one 
is even much more expensive. At this stage we are 
however only interested in understanding the efficiency of a certain type of move and we will for the moment not care about the 
cost of a single move. We will measure the efficiency of the moves by counting how many local minima will be visited on average 
in this modified saddle point search based version of the Minima Hopping algorithm before the global minimum is found 
and we will ignore the fact that the CPU 
time can be very long due to the cost of the moves. In our implementation of this method we perform 50 saddle point searches starting from the current local minimum. Out of these we choose the one exhibiting the lowest barrier. Since the saddle point searches sometimes give saddle points which are not connected to the initial minimum (meaning that a local geometry optimization starting at that saddle point would not lead back to the initial minimum), these barriers may be meaningless. However, we do not care about this fact and simply choose that saddle point with the lowest value.

An overview of the performance with this method, which we denote as lowest barrier (LB), is given in Table~\ref{tab: comp38and55}. We compare the performance of the LB method to that of the standard molecular dynamics (MD) version and to that of the saddle point based lowest mode (LM) method which will be discussed in section \ref{Following the lowest mode}.  As usual~\cite{sandro} we 
start our molecular dynamics trajectory in a soft direction, i.e. in a direction with low curvature in order to overcome 
a low barrier with a small number of molecular dynamics steps.  This direction should however not exactly be identical to 
the direction of the lowest curvature, i.e. the lowest eigenvector of the Hessian matrix, because we would again loose ergodicity 
in this way. We need enough randomness in the initial direction of the velocity vector to be able to jump into different 
catchment basins when we escape repeatedly from a certain minimum.

As one sees, the LB method is somewhat more efficient than MD in terms of the number of distinct local minima that are visited 
before finding the global minimum for $LJ_{38}$. In terms of the total number of minima the MD based escapes are however more efficient. 
This comes from the fact that the MD based escapes are more ergodic than the LB based escapes and repeated visits of the 
same minimum are therefore less likely. Fig.~\ref{fig: curvatureBarrier.eps} shows that a MD trajectory has 
a large choice for crossing very low barriers, i.e. there are many low lying saddle point around a local minimum. In the case of $LJ_{55}$ the MD based escapes are more efficient according to both criteria.
The surprising result that MD is more efficient than LB comes from the fact that the molecular dynamics trajectory can 
cross several barriers whereas in the LB based moves one crosses by definition only one barrier. In the case of the $LJ_{55}$ 
cluster crossings of several barriers are frequently encountered since the whole energy landscape is strongly 'tilted' in the direction 
of the global minimum. 


\begin{table}[h]
\centering
\begin{tabular}{l r r m{.25cm} r r}
\hline
& \multicolumn{2}{c}{\small $LJ_{38}$} & & \multicolumn{2}{c}{\small $LJ_{55}$} \\
 \cline{1-3} \cline{5-6}
  & \multicolumn{1}{p{1.3cm}}{\small \centering  total minima} & \multicolumn{1}{p{1.3cm}}{\small \centering different minima} & & \multicolumn{1}{p{1.3cm}} {\small\centering total minima} & \multicolumn{1}{p{1.3cm}} {\small\centering different minima} \\
{\small SP LM} & {\small  $2703.6$} & {\small  $523.9$} & & {\small  $415.5$} & {\small  $91.9$} \\
{\small MD} & {\small $1030.1$} &  {\small $297.4$} & & {\small  $92.3$} & {\small \centering $28.4$} \\
{\small SP LB} & {\small $1626.1$} & {\small  $268.2$} & & {\small  $584.0$} & {\small  $96.3$} \\
\hline
\end{tabular}
\caption{Comparison of the performance of all three Minima Hopping versions for both $LJ_{38}$ and $LJ_{55}$ clusters. ``total minima'' are the total number of visited minima per run, ``different minima'' indicates how many among them are different. ``SP LM'' means the saddle point version that follows the lowest mode, ``SP LB'' the one that crosses the lowest barrier. The data is based on 100 runs for each version.} 
\label{tab: comp38and55}
\end{table}

\subsection{Following the lowest mode}
\label{Following the lowest mode}
\begin{figure}[h]
 \includegraphics[width=.45\textwidth]{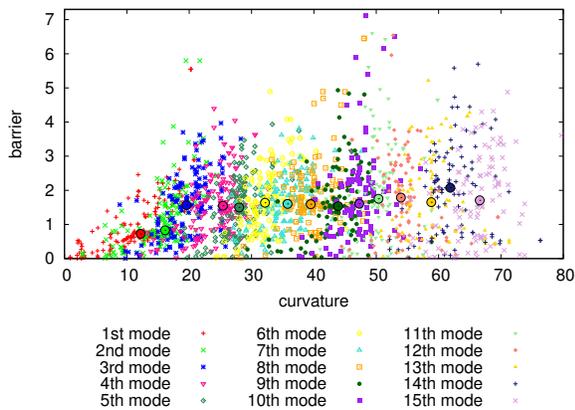}
\caption{A plot of the barrier heights as a function of the curvature along the direction in which 
the saddle point search was started. In addition the numbers of the modes along which the search 
was started are distinguished by colours. The median
value for each mode (median value of the curvature 
as well as the median value of the barrier) is plotted with a large dot. One can see that it is very unlikely to 
find high barriers along the softest (lowest curvature) directions. Along the stiffer (higher curvature) 
directions one finds increasingly higher barriers 
but in addition there exist also low barriers. }
\label{fig: curvatureBarrier.eps}
\end{figure}

Even though the results for the LB case are already
discouraging, we present a second scheme which is
more realistic 
since it does not require to find all the barriers around a local minimum to make a single move. 
In this scheme we exploit the fact that there is a correlation 
between the curvature of the direction into which we start our saddle point search and the height of the saddle point found.
This correlation is shown in Fig.~\ref{fig: curvatureBarrier.eps} and it is the same correlation that is also exploited when we start our molecular dynamics 
trajectories in soft directions. In this scheme, which we denote by lowest mode (LM), we search for the saddle point 
in the softest direction if the minimum is visited for the first time, in the second softest direction 
if it is visited for the second time and so on. Using DIIS for the dimer rotation ensures that these searches will lead to distinct saddle points with high probability. If we find a saddle point we will move the system over this saddle point and 
perform a local geometry optimization. Each mode gives us two degenerate directions, namely the direction of the 
positive and negative eigenvector of this mode, and we can therefore perform two saddle point searches for each mode.

\begin{figure}[h]
 \includegraphics[width=.45\textwidth]{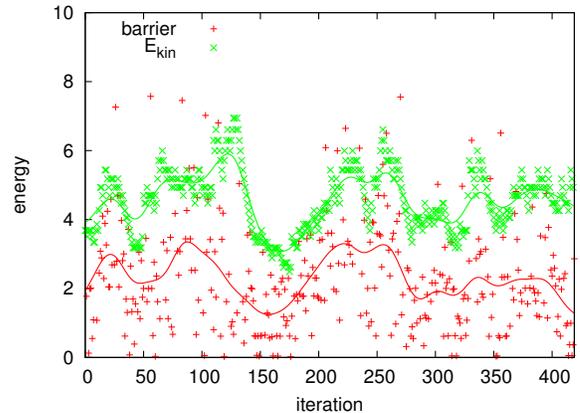}
\caption{The correlation between barrier heights and kinetic energy during a MD Minima Hopping run. 
The kinetic energy is scaled down by a factor of three
for a better visualization. The straight lines are convolutions of the data with a Gaussian.
There is a clear correlation between the barrier height and the kinetic energy, even if the kinetic energy
is much higher than the barriers.}
\label{fig: barriersMDMinHop.eps}
\end{figure}

The results in Table~\ref{tab: comp38and55} show that this approach is in all cases much less 
efficient than the molecular dynamics based moves. It is due to the fact that even if we start our saddle point search in a soft 
direction we can frequently obtain very high saddle points, whereas in the molecular dynamics based moves energy conservation 
will prevent the crossing of such high barriers. The energy of the molecular dynamics trajectory is usually much larger than needed to cross a barrier, and one would therefore not expect that there is a correlation between the energy of the trajectory crossing from one catchment basin into another one and the height of the saddle point that connects the minima of the two 
catchment basins. However, we found that such a correlation does indeed exist, as shown in Fig~\ref{fig: barriersMDMinHop.eps}.


\section{Moves for binary systems}
In the case of binary systems one might think that molecular dynamics based moves are inefficient since atoms of a certain type 
will only move by some slow diffusive motion to their right place. A process which can bring atoms potentially faster to their 
right place is an identity exchange where the identity of two atoms that are possibly far apart is exchanged. In a binary 
Lennard-Jones (BLJ) cluster the identities will be denoted by A and B. On the other hand we know that the efficiency of the molecular 
dynamics based moves in Minima Hopping is due to the fact that high energy configurations are rarely visited which leads to 
small values of $E_{diff}$. Table~\ref{ediff} shows these values for several BLJ systems which have different size-mismatch values $\sigma$.
The $E_{diff}$ value is always the value which gives on average an acceptance ratio of 0.5 in the standard MD based version of MH.
$\sigma = \frac{\sigma_{BB}}{\sigma_{AA}} $ is the size of the larger type-B atom in a BLJ-cluster where the smaller type-A atoms are chosen to have size 1. The 
interaction potential is then given by
$$  E=4 \sum_{i<j} \epsilon_{\alpha\beta} \left[\left(\frac{\sigma_{\alpha\beta}}{r_{ij}}\right) ^{12}-\left(\frac{\sigma_{\alpha\beta}}{r_{ij}}\right)^6\right]    $$
where $\alpha$ and $\beta$ are the types of atoms $i$ and $j$. $\sqrt[6]{2} \sigma_{\alpha \beta}$ 
is the equilibrium pair separation  and $\epsilon_{\alpha\beta}$
is the well depth of the pair potential from atoms  $i$ and $j$. We set
 $ \epsilon_{AA}=\epsilon_{BB}=\epsilon_{AB}=\epsilon=1 $ and
$ \sigma_{AB}=\frac{\sigma_{AA}+\sigma_{BB}}{2} $. With these settings, the only free parameter besides the number
 of type-A atoms $ N_A $ is  $\sigma$ which is chosen to be $ \sigma \in \{1,1.05,...,1.3\}$.

\begin{table}[h]
\centering
\begin{tabular}{llll}\\\hline
\hspace*{2mm}$\sigma$ & system & $E_{diff}$ & acc./rej. \\\hline
1.05\hspace{5mm} & $A_{19}B_{26}$\hspace{6mm} & \hspace{1mm}0.44\hspace{5mm} & \hspace{2mm} 1.50  \\
1.15\hspace{5mm} & $A_{31}B_{63}$\hspace{6mm} & \hspace{1mm}0.66\hspace{5mm} & \hspace{2mm} 0.14  \\
1.20 \hspace{5mm} & $A_{33}B_{63}$\hspace{6mm} & \hspace{1mm}0.37\hspace{5mm} & \hspace{2mm} 0.03  \\
1.25\hspace{5mm} & $A_{42}B_{58}$\hspace{6mm} & \hspace{1mm}0.46\hspace{5mm} & \hspace{2mm} 0.01  \\\hline
\end{tabular}
\caption{acceptance/rejectance ratio for identity exchange moves with fixed $E_{diff}$.}
\label{ediff}
\end{table}

\begin{figure}[h]
 \includegraphics[width=.45\textwidth]{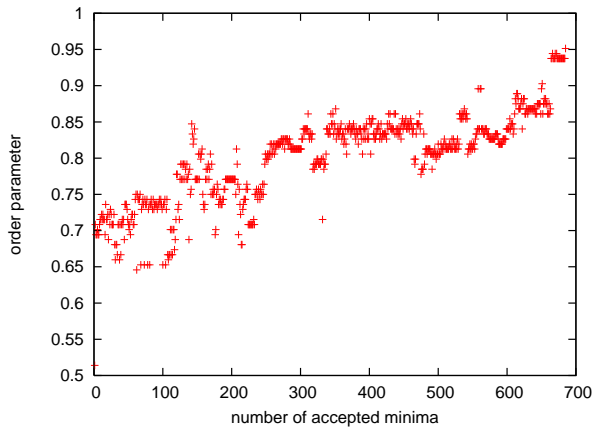}
\caption{Figure shows how fast a core shell structure of a BLJ-cluster is formed if one starts from 
 a random distribution of atoms of type A and B. The order parameter is defined as the fraction of atoms which 
are in the core. An order parameter of 1 corresponds thus to the global minimum structure of a particular system. The evolution towards 
the core-shell structure is plotted only until the
order parameter reaches the value 0.95 where 
the core-shell structure is already clearly visible. The data is obtained for the system $A_{39}B_{12}$ 
with $\sigma=1.1$ }
\label{CSevo}
\end{figure}

Table~\ref{ediff} also shows the acceptance ratio for identity exchange moves followed by a local geometry optimization, 
if the $E_{diff}$ of the MD move for the same system is used. One can see that 
these acceptance ratios get smaller and smaller with increasing $\sigma$. This means that 
in most cases exchanging two atom types will lead to rather high energy configurations and is hence less efficient than 
MD moves. In nature real atoms do not only differ by size (e.g. covalent radius) but also by their 
electronic properties.  Exchange moves are therefore expected to be efficient only if the atoms are very similar in 
every respect. 
If the atoms are very different there is actually a strong driving force present in the MD moves to 
put the different types of atoms at the right positions. If the global minimum structure is for instance a core-shell 
structure we obtain a core-shell like structure starting from a random position already after some 100-1000 MD moves 
(see Fig~\ref{CSevo}). Atomic identity exchange moves are therefore not only not necessary, but would even 
be counterproductive. 

For systems with two similar types of atoms (i.e. small  $\sigma$ for BLJ-systems) identity exchange moves can 
however reduce the average search time for the global minimum. Due to the significantly lower acceptance ratio 
at constant $E_{diff}$, these exchange moves can not be treated on an equal footing as the MD moves. In particular it would be too 
expensive to do a local geometry optimization after each exchange move. For this reason we have incorporated exchange moves 
in the following way in our Minima Hopping algorithm: After each MD move we do a number of exchange moves which is roughly 
equal to the number of force evaluations required in the geometry optimization. If the energy of the unrelaxed configuration 
resulting from this exchange move increases by less than $E_{relax}$ with respect to the original configuration it is relaxed 
and taken as the result of this combined MD/exchange move. $E_{relax}$ is the energy that is on average gained by 
a local geometry optimization starting from a relaxed configuration where the identity of two atoms is exchanged. 
Hence the relaxed energy of the exchanged configuration will be on average lower then the energy of the original 
configuration. 
In this way the exchange moves can help in finding the global 
minimum even if their acceptance probability is lower than the acceptance probability of the MD moves by a factor which is 
roughly equal to the number of force evaluations needed by the MD moves. 

\section{Results for the binary LJ benchmark systems}
Finding the global minimum for a binary Lennard-Jones cluster is significantly more difficult than for a mono-atomic
Lennard-Jones cluster. 
In the limit where the two atoms become identical (but are still slightly different) each configuration becomes degenerate 
with a degeneracy of $\frac{ (N_A + N_B)! }{ N_A! N_B!}$, where $N_A$ is the number of BLJ atoms of type A and $N_B$ the 
number BLJ atoms of type B. Such configurations, that can be transformed into each other by identity exchanges of the atom types,  
are called homotopes~\cite{homotope}.

If the type-A and type-B atoms have a larger
size-mismatch $\sigma$, 
not all $\frac{ (N_A + N_B)! }{ N_A! N_B!}$ homotopes
are stable, but nevertheless a smaller number of 
homotopes exist. The existence of stable
homotopes increases the number of local minima of
binary systems compared to the mono-atomic case.
Depending on the system size and composition this
number can thus be significantly larger.  
The difficulty of the global optimization of binary
Lennard-Jones systems is also reflected in the data of
the Cambridge Cluster database~\cite{database}. 
For mono-atomic Lennard-Jones systems the putative
global minima up to a cluster size of 1000 atoms are
listed, but for 
BLJ-systems only up to 100 atoms.  Since the putative global minima structures are given for 
6 different size ratios $\sigma$ the database contains 600 structures. 
A first computation
 of the putative global minima in this database was done by Doye \textit{et al.}~\cite{doye_2005}.  
Andrea Cassioli, Marco Locatelli and Fabio Schoen~\cite{fabio} 
reexamined the problem and found nearly 100 new putative global minima for 
the 600 structures in the database. A few new structures were also found by Pullan~\cite{pullan}. 
In spite of the fact that several groups 
have already reexamined the database we were able to find the following 17 structures which are 
lower in energy than the structures listed in the
database (status June 2006):

\begin{itemize}
\item $\sigma=1.30$: $BLJ_{100}$, 
$ BLJ_{99} $, $BLJ_{98}$, $BLJ_{97}$,
$BLJ_{96}$
\item $\sigma=1.25$: $BLJ_{100}$,  
$BLJ_{99}$, $BLJ_{98}$, $BLJ_{97}$, 
$BLJ_{96}$
\item $\sigma=1.20$: $BLJ_{100}$,
$BLJ_{97}$, $BLJ_{96}$, 
$BLJ_{95}$
\item $\sigma=1.15$: $BLJ_{94}$, $BLJ_{93}$, $BLJ_{92}$.
\end{itemize}

The global optimization runs for systems with a size-mismatch 
ratio of $\sigma\geq 1.2$ were done with the standard MH
algorithm \cite{minhop}. The putative global minima
with $\sigma=1.15$ have been found by using the
above mentioned identity moves which turned out 
to be a powerful additional feature to the MH algorithm
for the global optimization of binary systems with
comparable atomic sizes.\\

We did not systematically recalculate the whole
Cambridge cluster database. 
However, a visualization of the putative global minimum
structures provided by the database revealed that
several structures didn't fit into the ``series'' under the 
same $\sigma$ due to 
too much disorder of the clusters or a still incomplete
separation of core and shell. 
These structures were reexamined and new energetically lower structures, 
that frequently had also different stoichiometric
compositions, were found in many cases. 

\subsection{New putative global minima}
\begin{figure}[!hbt]
\centering
\subfigure[]{ 
\includegraphics[width=0.15\textwidth]{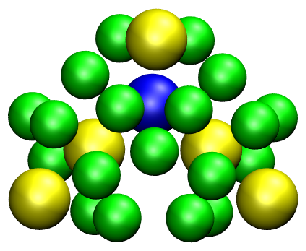}
            }
\subfigure[]{   
\includegraphics[width=0.13\textwidth]{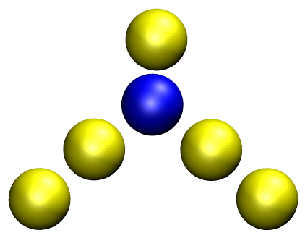}
            }
\subfigure[]{

\includegraphics[width=0.15\textwidth]{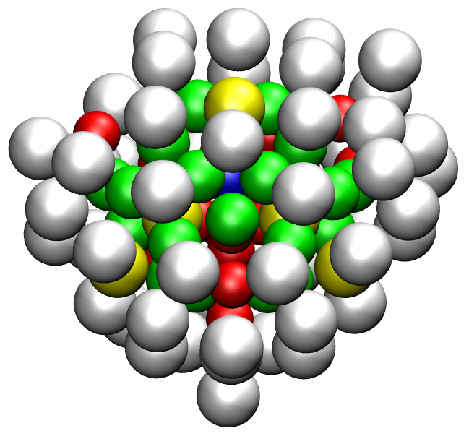} 
            }
 
\caption{Putative global
minimum structure of $BLJ_{97,\sigma=1.25}$ represented
by the $A_{39}B_{58}$ composition. a) 3
(single) disclination lines (yellow) passing edges with
6 tetrahedra around them (green), connected at the
$Z15$-node (blue). b) Disclination network
coloured. c) Shows the whole cluster in corresponding
colours}
  \label{clust97_39_125}  
\end{figure}

We now present the structures corresponding to the new
putative global minima. Using the classification 
criteria of Doye \textit{et al.}~\cite{doye_2006}, we will assign all cluster
structures to their structural type families
and their symmetry point groups.\\
A new class of structures which introduces a new region in the structural phase diagram \cite{doye_2006}, \cite{doye_2005}
for large system sizes $N \geq 98 $ and $\sigma \geq 1.25 $ is
the global minimum structure of $BLJ_{100,\sigma=1.3}$. The polytetrahedral structure 
with disclination network can be classified as $4Z14$ structure with point group symmetry
$C_{S}$. Generally, $Z14$-atoms 
are part of a single disclination line whereas
$Z15$ (see Fig.\ref{clust97_39_125}) and $Z16$ atoms act as nodes connecting 3 or 4
disclination lines respectively. Disclination lines always pass edges with six
tetrahedra around them, see Fig.\ref{clust97_39_125}\\
The 4 atoms with coordination number $Z=14$ form 4
pairwise disconnected (single) disclination lines ending 
at 4 $Z=13$ shell-atoms\cite{footnote}. This type of structure is
also the corresponding putative global minimum
structure
of $BLJ_{99,\sigma=1.3}$, $BLJ_{100,\sigma=1.25}$,
$BLJ_{99,\sigma=1.25}$ and $BLJ_{98,\sigma=1.25}$.
The core of these structures consists of 42 atoms and
is completely covered by the shell atoms (pure core-shell).
Depending on the cluster size there is only an absence
of one or two type-B shell atoms. The putative ground
state energy of $BLJ_{100,\sigma=1.3}$  is \mbox{-604.796307}
in common units.

\begin{figure}[!hbt]
\centering
\subfigure[]{   
\includegraphics[width=0.14\textwidth]{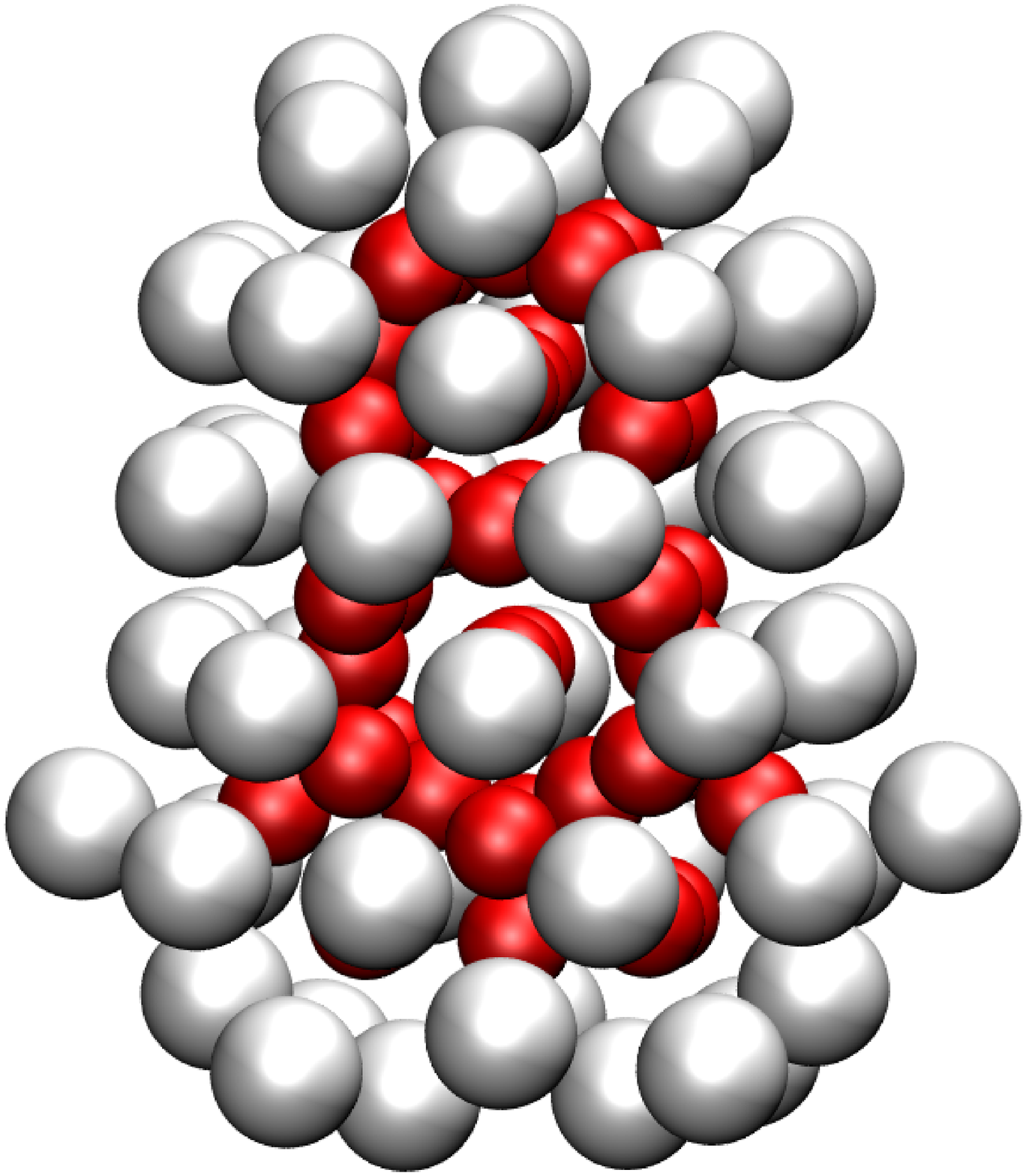}
            }
\subfigure[]{   
\includegraphics[width=0.14\textwidth]{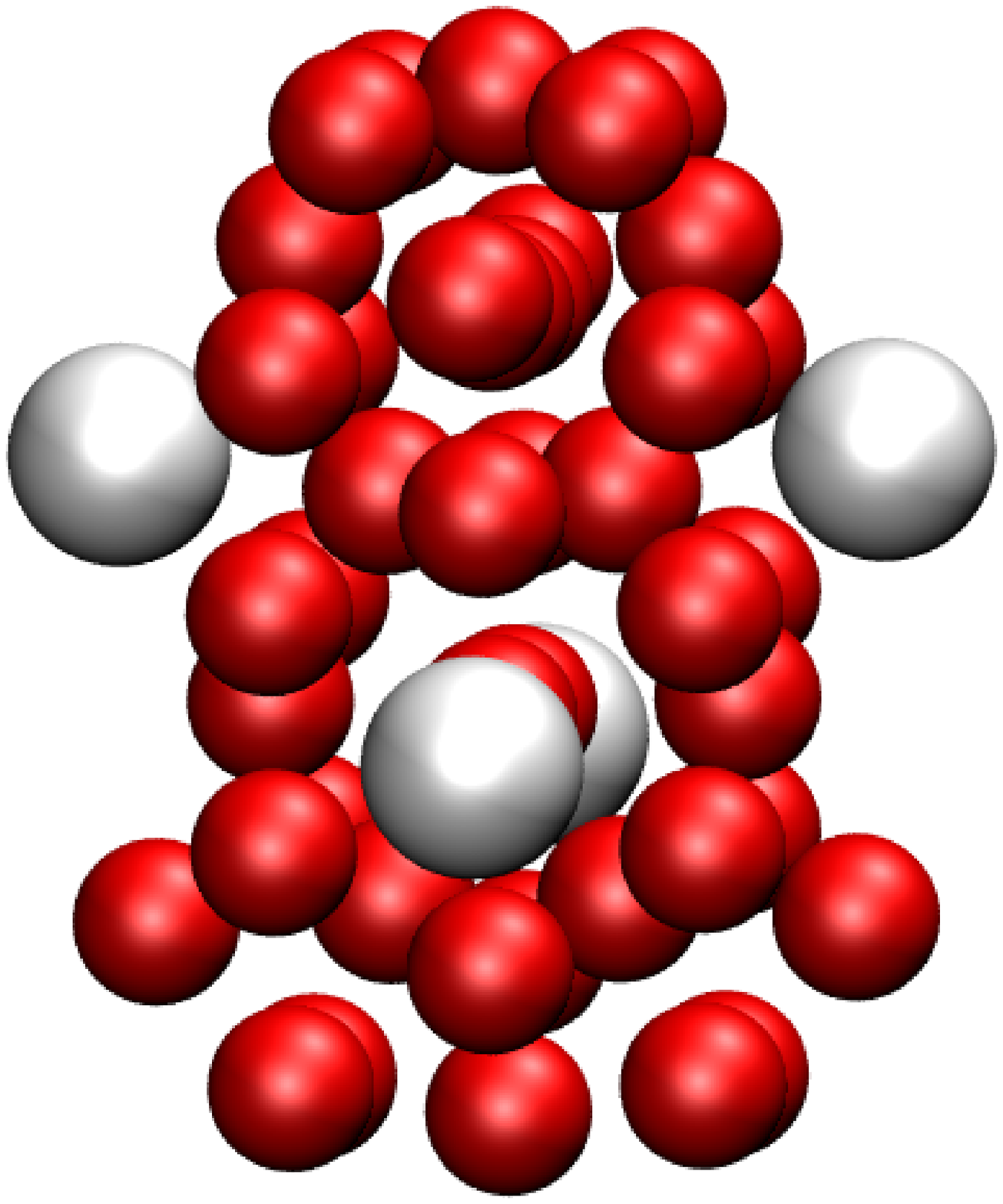}
  }
\subfigure[]{    
\includegraphics[width=0.125\textwidth]{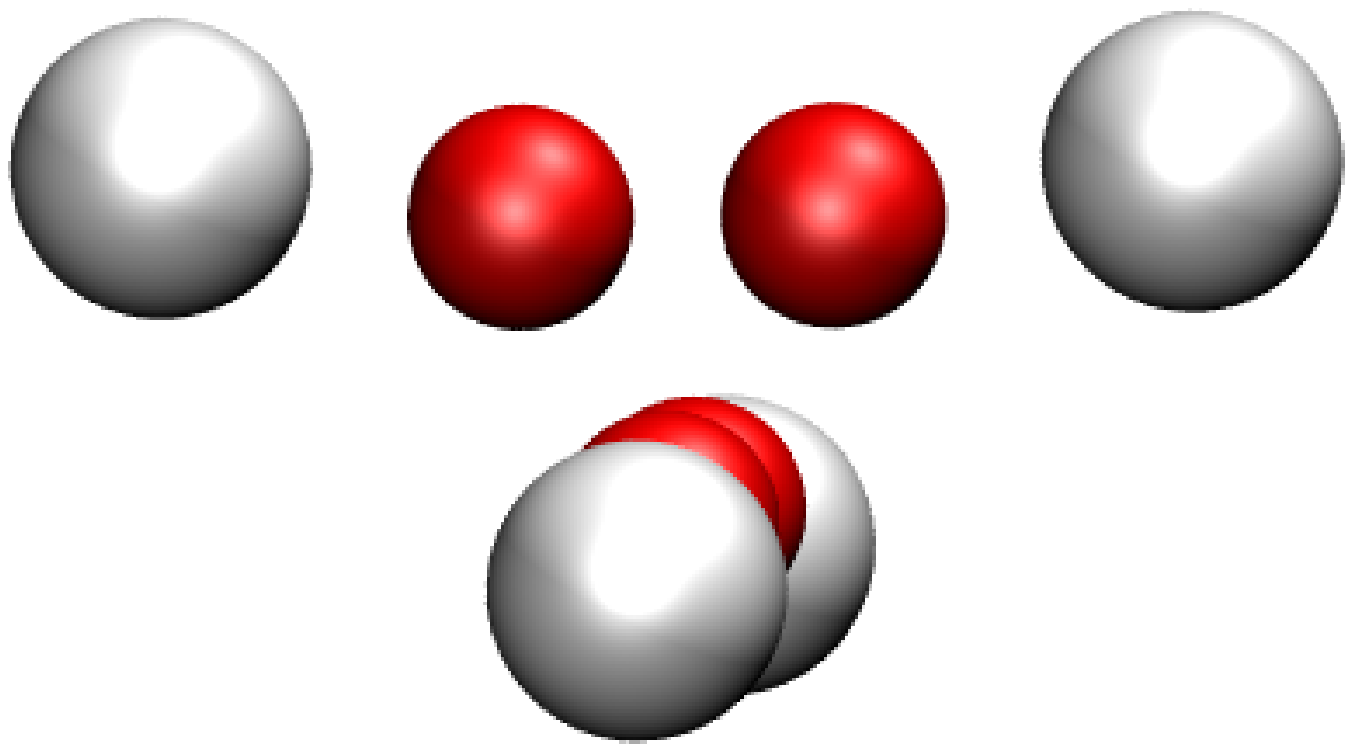} 
             }

\caption{$4Z14$ structure of $BLJ_{100,\sigma=1.3}$
with $C_{S}$ symmetry and the putative ground state
energy -604.796307. a) Shows the whole cluster.  b)
Disclination network embedded into the core. c)
Disclination network}
  \label{clust100_42_13}  
\end{figure}
A summary of the structural types and energies corresponding to all other minima we found is given
in table \ref{summarytable}. 
Fig. \ref{clust97_39_125} shows the
$A_{39}B_{58}$ composition of
$BLJ_{97,\sigma=1.25}$ and how the disclinations are
embedded into the whole cluster. It is a polytetrahedral
$Z15$ structure with $C_{1}$ symmetry and the typical $Z15$-disclination network. The putative global minimum energy is \mbox{-578.201634}.

Doye \textit{et al.}~\cite{doye_2006} describe the
structural motives of other common binary Lennard-Jones
structures up to a system size $N=100$.\\

\begin{table}[h]
\centering
\begin{center}
\begin{tabular}{llllll}\hline
   \hspace{1mm}$N$   &\hspace{3mm} $\sigma$ &\hspace{8mm} $\epsilon$ &\hspace{1mm} stoichio- &\hspace{1mm} point  &\hspace{1mm} structural \\ 
   \hspace{1mm} &\hspace{1mm}          &\hspace{1mm}            &\hspace{2mm} metry     &\hspace{1mm} group  &\hspace{5mm}  type \\ \hline 
100 \hspace{1mm}&\hspace{1mm} 1.3      &\hspace{1mm} -604.796307 &\hspace{1mm}   $A_{42}B_{58}$ &\hspace{3mm} $C_{s}$ & \hspace{4mm}4Z14 \\ 
99  \hspace{1mm}&\hspace{1mm} 1.3      &\hspace{1mm} -597.592233 &\hspace{1mm}   $A_{42}B_{57}$ &\hspace{3mm} $C_{s}$ & \hspace{4mm}4Z14 \\ 
98 \hspace{1mm} &\hspace{1mm} 1.3      &\hspace{1mm} -590.787413 &\hspace{1mm}   $A_{41}B_{57}$ &\hspace{3mm} $C_{1}$ &   \hspace{4mm}Z14 \\ 
97  \hspace{1mm}&\hspace{1mm} 1.3       &\hspace{1mm} -583.871531 &\hspace{1mm}  $A_{41}B_{56}$ &\hspace{3mm} $C_{1}$ &   \hspace{4mm}Z14 \\ 
96  \hspace{1mm}&\hspace{1mm} 1.3       &\hspace{1mm} -576.517002 &\hspace{1mm}  $A_{41}B_{55}$ &\hspace{3mm} $C_{1}$ &   \hspace{4mm}Z14   \\ 
100 \hspace{1mm}&\hspace{1mm} 1.25     &\hspace{1mm} -599.264624 &\hspace{1mm}   $A_{42}B_{58}$ &\hspace{3mm} $C_{s}$ & \hspace{4mm}4Z14 \\ 
99  \hspace{1mm}&\hspace{1mm} 1.25      &\hspace{1mm} -592.138846 &\hspace{1mm}  $A_{42}B_{57}$ &\hspace{3mm} $C_{s}$ & \hspace{4mm}4Z14 \\ 
98 \hspace{1mm} &\hspace{1mm} 1.25      &\hspace{1mm} -584.930661 &\hspace{1mm}  $A_{42}B_{56}$ &\hspace{3mm} $C_{s}$ & \hspace{4mm}4Z14 \\ 
97 \hspace{1mm} &\hspace{1mm} 1.25      &\hspace{1mm} -578.201634 &\hspace{1mm}  $A_{39}B_{58}$ &\hspace{3mm} $C_{1}$ &   \hspace{4mm}Z15 \\ 
96  \hspace{1mm}&\hspace{1mm} 1.25      &\hspace{1mm} -571.389275 &\hspace{1mm}  $A_{39}B_{57}$ &\hspace{3mm} $C_{1}$ &   \hspace{4mm}Z15 \\ 
100 \hspace{1mm}&\hspace{1mm} 1.2     &\hspace{1mm} -591.768143 &\hspace{1mm}    $A_{35}B_{65}$ &\hspace{3mm} $C_{1}$ & \hspace{4mm}Z16Z15 \\ 
97 \hspace{1mm} &\hspace{1mm} 1.2     &\hspace{1mm} -571.392434 &\hspace{1mm}    $A_{33}B_{64}$ &\hspace{3mm} $C_{1}$ & \hspace{4mm}Z16Z15 \\ 
96 \hspace{1mm} &\hspace{1mm} 1.2     &\hspace{1mm} -564.674461 &\hspace{1mm}    $A_{33}B_{63}$ &\hspace{3mm} $C_{1}$ & \hspace{4mm}Z16Z15 \\ 
95 \hspace{1mm} &\hspace{1mm} 1.2     &\hspace{1mm} -557.690639 &\hspace{1mm}    $A_{33}B_{62}$ &\hspace{3mm} $C_{1}$ & \hspace{4mm}Z16Z15 \\ 
94 \hspace{1mm} &\hspace{1mm} 1.15    &\hspace{1mm} -542.476905 &\hspace{1mm}    $A_{31}B_{63}$ &\hspace{3mm} $C_{1}$ & \hspace{4mm}Z16Z15 \\ 
93 \hspace{1mm} &\hspace{1mm} 1.15    &\hspace{1mm} -535.594853 &\hspace{1mm}    $A_{31}B_{62}$ &\hspace{3mm} $C_{1}$ & \hspace{4mm}Z16Z15 \\ 
92 \hspace{1mm} &\hspace{1mm} 1.15    &\hspace{1mm} -529.190149 &\hspace{1mm}    $A_{28}B_{64}$ &\hspace{3mm} $C_{1}$ & \hspace{4mm}Z16Z15 \\\hline
\end{tabular}
\end{center}
\caption{ System size $N$, energies $\epsilon$, compositions, point groups and structural types of the new putative global minima.}
\label{summarytable}
\end{table}

\section{Conclusions}
Molecular dynamics based moves were found to be optimal in the context of global optimization. 
According to the Bell Evans Polyani principle one should escape over low barriers. 
Even though moves that escape over the lowest saddle point 
around a local minima are consequently expected to perform better, if the number of visited minima is taken as the success 
criterion, it turns out that molecular dynamics based moves are more efficient for energy landscapes which have a 
strong funnel like  
 structure, since in this case the MD trajectory can cross over several saddle points. 
For other energy landscapes MD based escapes are about equally efficient in terms of the number of distinct visited minima, 
but are more efficient in terms of the total number of visited minima, i.e. in terms of the number of 
local geometry optimizations. 
In practice moves that escape over the lowest saddle point are too expensive since they require a complete 
exploration of the potential energy surface around the local minimum from which one wants to escape. 
In practice the only saddle point escape moves that are affordable would be moves where one searches for a single 
saddle point in a soft direction. It turns out that these saddle points can sometimes be very high and 
the approach is therefore much less efficient that an approach using molecular dynamics 
 based moves 
in soft directions. The important difference is that by energy  conservation the MD based trajectory can not 
cross over energetically high saddle points whereas saddle point searches in soft directions can 
give very high saddle points. In addition MD based escapes require significantly less force evaluations 
(of the order of 100) than even a single saddle point
search. 
The value of the parameter $E_{diff}$ in the Minima
Hopping method is a good measure of the quality of the
moves.
If moves lead on average in other low energy configurations $E_{diff}$ will be small and one has to search only over 
low energy structures. One has therefore to search only over a number of local minima which is much smaller than 
in the case where one has to search in a larger energy window above the global minimum. According to this criterion 
identity exchange moves are in general worse than MD based moves except for very small values of $\sigma$. If 
identity exchange moves are however added as some kind of post processing step to a MD based move without the need 
of an additional geometry optimization for each exchange trial, the efficiency of the Minima Hopping method can be improved.  
With such an improved version of the Minima Hopping method we were able to find several new global minima structures for 
binary Lennard Jones clusters with up to 100 atoms and size ratios of $\sigma=1.15$. For large values of $\sigma$ the 
ordinary Minima Hopping method without identity exchanges was used and  turned out to be powerful enough to 
find new global minima structures for size rations of
$\sigma =$ 1.2, 1.25 and 1.3.

Financial support from SNF and computing time from CSCS are acknowledged. 
We gratefully acknowledge expert discussions with Riccardo Ferrando.

\end{document}